\documentclass[a4paper,11pt]{article}
\usepackage{pos}
\usepackage{mathbbol}  

\newcommand{\sun}{$\mathrm{SU}(N)$}
\newcommand{\sut}{$\mathrm{SU}(3)$}
\newcommand{\sutw}{$\mathrm{SU}(2)$}
\newcommand{\Nc}{N_\mathrm{C}}

\title{Normalizing flows for \sun~gauge theories employing singular value decomposition}

\author*[a]{Javad Komijani}
\author[a]{Marina K. Marinkovic}

\affiliation[a]{Institute for Theoretical Physics, ETH Zurich, 8093 Zurich, Switzerland}

\emailAdd{jkomijani@phys.ethz.ch, marinama@ethz.ch}

\abstract{%
We present a progress report on the use of normalizing flows for generating gauge  
field configurations in pure \sun~gauge theories. We discuss how the singular value  
decomposition can be used to construct gauge-invariant quantities, which serve as
the building blocks for designing gauge-equivariant transformations of \sun~gauge links. Using this novel approach, we build  
representative models for the \sut~Wilson action on a \(4^4\) lattice with  
$\beta = 1$. We train these models and provide an analysis of their performance,  
highlighting the effectiveness of the new technique for gauge-invariant  
transformations.
We also provide a comparison between the efficiency of the proposed algorithm and the spectral flow of Wilson loops.
}

\FullConference{The 41st International Symposium on Lattice Field Theory
(LATTICE2024)\\
28 July - 3 August 2024\\
Liverpool, UK\\}

\begin{document}
\maketitle

\section{Introduction}
\label{sec:intro}

The method of normalizing flows (NF) is a generative approach for sampling 
from complex probability distributions. The samples are generated by applying 
a series of invertible transformations to samples drawn from a simpler, prior 
distribution. In the area of lattice field theory, this method has been applied 
to several models, including gauge theories; see, e.g., 
Refs.~\cite{Kanwar:2020xzo, Boyda:2020hsi, Kanwar:2024ujc, Bacchio:2022vje}.

In this work, we apply NF on the Wilson's discretization of the \sut~gauge theory. 
We construct gauge-equivariant transformations by applying singular value 
decomposition (SVD) to the sum of adjacent staples of each link. By imposing
gauge-equivariance, the NF transformation becomes aligned
with the structure of the underlying theory, consequently improving the training.

As the prior, we use \sut~links generated uniformly with respect to the Haar measure.
We expect the model to achieve higher efficiency if the prior
resembles the target distribution. 
To address this, we also apply the transformations to an alternative prior
inspired by trivializing maps~\cite{Luscher:2009eq}.
A trivializing map is itself a continuous normalizing flow (CNF), which provides a 
natural bridge between simpler priors and more complex ones.

Here, we first briefly discuss trivializing maps and apply their 
leading-order approximation to the \sut~gauge theory with the Wilson action on 
a $4^4$ lattice at $\beta=1$. We then propose how 
to use SVD to construct gauge-equivariant normalizing flows for \sun~gauge 
links with a focus on \sut. We construct an SVD-based representative model, 
train it and its variants---including two different priors---and discuss
their training efficiency.

\section{Trivializing maps and Wilson flow}
\label{sec:Wilson-flow}

A CNF consists of infinitesimal transformations and is described as an 
ordinary differential equation with respect to a fictitious flow time \( t \). 
With a similar premise, Ref.~\cite{Luscher:2009eq} introduces the method of 
trivializing maps for gauge theories as:
\begin{equation}
   \frac{d}{dt} V_t = f(V_t, t) V_t.
   \label{eq:triv-map}
\end{equation}
Here, $V_t$ at $t = 0$ and $t = 1$ represents \sun~link  variables
corresponding to the prior and target distributions, respectively%
\footnote{In Ref.~\cite{Luscher:2009eq}, the flow is in the reverse direction 
of the convention adopted here.}.
For the Wilson gauge action,
\begin{equation}
    S_\text{W}[U] =
    -\frac{\beta}{2 \Nc} \sum_{x \in \Lambda} \sum_{\mu \neq \nu}
    \mathbb{Tr}~ U_\mu(x) U_\nu(x + \hat \mu) U_\mu^\dagger (x + \hat\nu)U_\nu^\dagger (x)
    \,,
    \label{eq:Wilson:Plaq}
\end{equation}
%
the leading-term trivializing map 
coincides with the Wilson flow up to an overall scaling factor:
\begin{equation}
   \frac{d}{dt } V_t = - \frac{\Nc}{2 (\Nc\,^2 - 1)} \frac{\beta}{2 \Nc} 
   \mathcal{P} \left\{V_t \Gamma_t \right\}\, V_t,
   \label{eq:triv-map}
\end{equation}
where $\Gamma_t$ is the sum of adjacent staples,
and $\mathcal{P}$ is a projection operator to the anti-hermitian traceless
space. The Jacobian of the transformation $V(0) \to V(t)$ reads
\begin{equation}
    \log J(t) = - \int_0^t d\tau\, S_\text{W}[V_\tau]\,.
\end{equation}

\begin{figure}
    \begin{center}
    \includegraphics[trim=3cm 0 4.5cm 0, clip, width=0.6\textwidth]{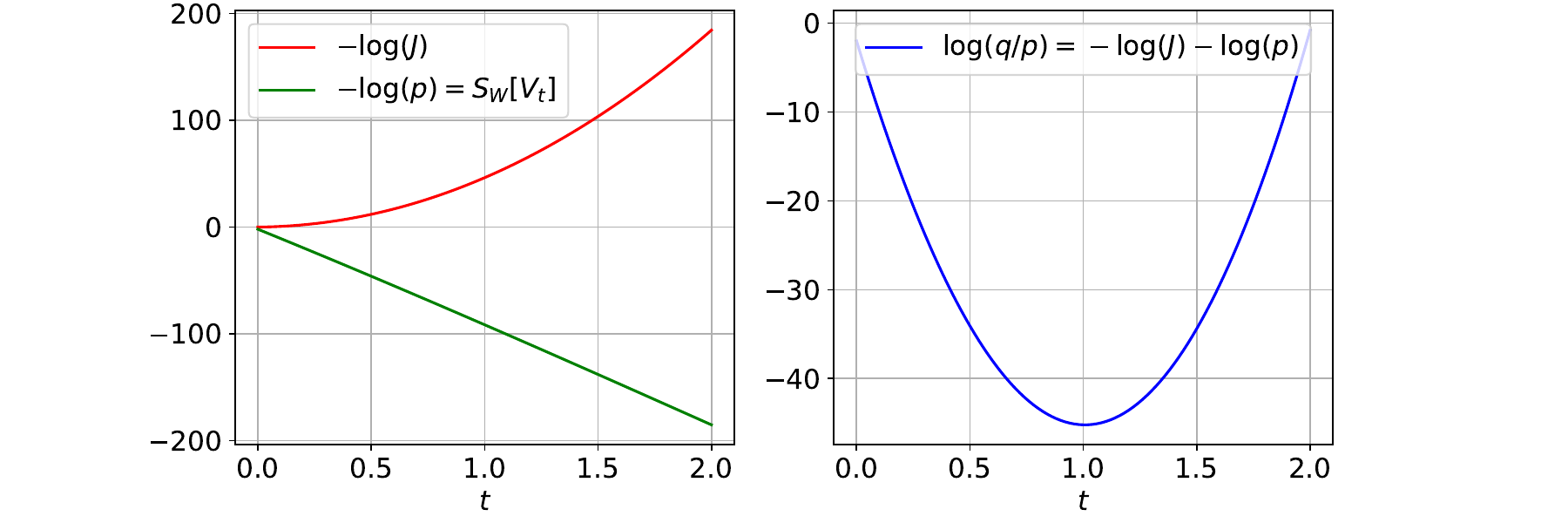}
    \end{center}
    \vspace{-0.5cm}
    \caption{Evolution of \sut~gauge links due to the (scaled) Wilson flow
    \eqref{eq:triv-map}.}
    \label{fig:LO_TM}
\end{figure}

Figure~\ref{fig:LO_TM} illustrates the flow applied to one configuration of
\sut~gauge links  on a \( 4^4 \) lattice with $\beta = 1$. The initial links
at $t = 0$ are uniformly drawn with the Haar measure. The left panel of 
Fig.~\ref{fig:LO_TM} shows the evolution of the Wilson gauge action, labeled as 
$-\log(p)$, along with the negative log-Jacobian of the transformation. The 
Wilson gauge action decreases monotonically, while the negative log-Jacobian
increases as the fictitious time increases. As shown in the right panel, the sum of these two
quantities reaches its minimum value at $t \approx 1$. This sum corresponds
to the Kullback-Leibler (KL) divergence, as discussed below.

The flow in equation \eqref{eq:triv-map} does not involve any trainable parameters. 
The flow from $t = 0$ to $t = 1$ transforms the uniform prior distribution of \sut~
gauge links into a distribution that more closely resembles the target distribution, 
with $p \propto e^{-S_\text{W}}$. In principle, higher-order terms can be included in
the trivializing map to further improve the similarity between the output of the flow
and the target distribution, as discussed in Refs.~\cite{Luscher:2009eq, Bacchio:2022vje}.
However, in this work, we do not consider such higher-order terms.
Instead, we use the flow to transform the prior distribution before feeding
it to NF-based models that we introduce below.
In fact, the Wilson flow serves a dual purpose in this work:
as a benchmark for evaluating the efficiency of our normalizing flow models
and as a tool for creating a prior distribution that resembles the target.

\section{Normalizing flows for gauge theories and gauge equivariance}

\begin{figure}
    \begin{center}
    \includegraphics[width=14cm]{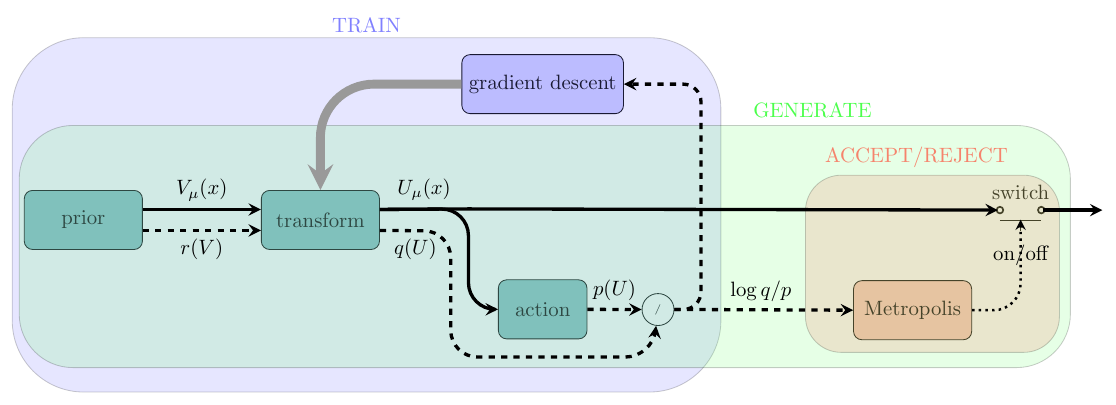}
    \end{center}
    \vspace{-0.5cm}
    \caption{Block diagram for the method of normalizing flows.
    $V_\mu(x)$ and $U_\mu(x)$ are the prior and transformed fields,
    and $r(V)$ and $q(U)$ are the corresponding probability densities.
    See Ref.~\cite{Komijani:2023fzy} for more information.}
    \label{fig:block-diag}
\end{figure}

Let us briefly highlight the essential components of the normalizing flows method. 
Three key elements are required: a prior distribution to draw initial samples, an 
invertible map with trainable parameters to transform the samples, and an action 
that defines the target distribution as $p \propto e^{-S}$. 
The NF model is then trained by minimizing the Kullback-Leibler (KL) divergence between 
the probability density function (PDF) of the transformed samples, denoted by $q$, 
and the target distribution $p$:
\begin{align}
   D_\text{KL}(q \parallel p) &\equiv \int dU ~ q(U) \log \frac{q(U)}{p(U)}
   ~~ \geq ~0.
   \label{eq:KL}
\end{align}
Note that the lower limit of the integral may shift if $p$ and/or $q$ are known 
only up to an overall normalization constant. Figure~\ref{fig:block-diag} illustrates 
the training process, which employs a self-learning method for optimizing the parameters.

In the context of gauge theories, the prior distribution of gauge links is  
typically chosen to be uniform with respect to the Haar measure. Since both  
the prior and target distributions inherently respect gauge symmetry, it is  
natural to employ gauge-equivariant transformations.
By explicitly incorporating gauge symmetry into these transformations, the  
model no longer needs to learn this symmetry, resulting in a more efficient  
training process. Additionally, this approach eliminates the risk of  
introducing unintended gauge fixing in the training process. 

A transformation is gauge-equivariant if it commutes with gauge transformations, 
thereby ensuring that gauge symmetry is preserved. One approach to construct such 
transformations is to work with gauge-invariant quantities. To this end, one can 
start by identifying the gauge-invariant quantities associated with the link variables. 
These quantities are then transformed, while the gauge-dependent components remain 
unchanged. Finally, the link variables are updated using the transformed gauge-invariant 
quantities and the frozen gauge-dependent components. To provide an example, we 
briefly review the gauge symmetry in the Wilson action for $\mathrm{SU}(\Nc)$ gauge theory 
and discuss the associated gauge-invariant quantities.

The Wilson action in equation \eqref{eq:Wilson:Plaq} is invariant under the 
following class of local transformations:
\begin{equation}
    U_\mu(x) \to Q(x)\, U_\mu(x)\, Q^\dagger(x + \hat\mu) ,
    \label{eq:gauge-transformation}
\end{equation}
where $Q(x)$ is an arbitrary \sun~matrix at each site of the lattice. 
The action depends on the links through the trace of the plaquette Wilson loops, 
\begin{equation}
    P_{\mu\nu}(x)
    = U_\mu(x) U_\nu(x + \hat \mu) U_\mu^\dagger(x + \hat\nu) U_\nu^\dagger(x) .
\end{equation}
More specifically, the action depends on the eigenvalues of $P_{\mu\nu}(x)$,
which are gauge invariant. 
Reference~\cite{Kanwar:2020xzo} uses these eigenvalues to construct
(plaquette-based) spectral flows.

The steps of a plaquette-based spectral flow for transforming the link variables $U_\mu(x)$ are
as follows. 
First, among all the plaquette Wilson loops containing $U_\mu(x)$, one is
chosen and labeled as $P_{\mu\nu}(x)$.
Then, the eigenvalues of $P_{\mu\nu}(x)$ are transformed to new values. 
Next, the updated plaquette Wilson loop, $P'_{\mu\nu}(x)$, is constructed using the 
updated eigenvalues and the unchanged eigenvectors. Finally, $U_\mu(x)$ is updated as
\begin{equation}
    U'_\mu(x) = P'_{\mu\nu}(x) P^\dagger_{\mu\nu}(x)  U_\mu(x) .
\end{equation}
To ensure the invertibility of the transformations, coupling layers are used with a 
masking pattern that divides all plaquette Wilson loops into frozen, active, and passive ones.

In the plaquette-based spectral flow described above, the link variable $U_\mu(x)$
is updated based on only one of the adjacent plaquette Wilson loops. However, when
$U_\mu(x)$ is updated, it (passively) affects other adjacent plaquette Wilson 
loops too. To avoid this passive updating, we propose a new method that 
updates the link variable $U_\mu(x)$ based on all adjacent plaquette Wilson loops.

By summing all the plaquette Wilson loops involving a designated link
$U_\mu(x)$, we express the Wilson action as
\begin{align}
    S_\text{W}[U] &= -\frac{\beta}{\Nc} 
    \mathbb{Re Tr} \left[ U_\mu(x) \Gamma_\mu(x)\right]
    + \text{rest}\,,
    \\
    \Gamma_{\mu}(x) &= \sum_{\nu \neq \mu} \Bigr\{U_\nu(x+\hat\mu) U^\dagger_\mu(x+\hat\nu) U^\dagger_\nu(x)
    - U_\nu^\dagger(x + \hat\mu - \hat\nu) U^\dagger_\mu(x-\hat\nu) U_\nu(x-\hat\nu)
    \Bigl\}
\end{align}
where $\Gamma_\mu(x)$ is the sum of all \emph{staples} corresponding to $U_\mu(x)$,
and ``rest'' contains for all other terms independent of $U_\mu(x)$.
We now apply singular value decomposition (SVD) to express $\Gamma_\mu(x)$ as
\begin{equation}
    \Gamma_\mu(x) = W_\mu(x) S_\mu(x) V^\dagger_\mu(x)\,,
\end{equation}
where $S$ is the diagonal matrix of singular values, and $W$ and $V$ are unitary matrices.%
\footnote{Note that $W$ and $V$ are not unique; for example, they can always 
be multiplied by a common overall phase factor.}
It turns out that the singular values are gauge invariant. Starting from the fact 
that under the gauge transformations \eqref{eq:gauge-transformation}, $\Gamma_\mu(x)$ 
transforms as
\begin{equation}
    \Gamma_\mu(x) \to Q(x + \hat\mu) \Gamma_\mu(x) Q^\dagger(x)\,,
\end{equation}
we conclude that the components of the SVD transform as
\begin{align}
  W_\mu(x) &\quad\to\quad Q(x+\hat\mu) W_\mu(x) \,,\\
  S_\mu(x) &\quad\to\quad S_\mu(x) \,,\\
  V_\mu(x) &\quad\to\quad V_\mu(x) Q^\dagger(x) \,.
\end{align}
Taking these transformations into account, we now perform a change of
variables as
\begin{equation}
  \tilde U_\mu(x) = V^\dagger_\mu(x) U_\mu(x) W_\mu(x) e^{-i\phi_\mu(x)}\,,
\end{equation}
where $e^{i\phi_\mu(x)}$ is a phase factor introduced to ensure that the 
\emph{transformed} link $\tilde U_\mu(x)$ remains special unitary, 
just like $U_\mu(x)$. One can easily verify that the transformed link
$\tilde U_\mu(x)$ is invariant under any gauge transformations.

Using the transformed link, we first express the Wilson gauge action as
\begin{align}
    S_\text{W}[U] &= -\frac{\beta}{\Nc} 
    \mathbb{ReTr}\left[ \tilde U_\mu(x) S_\mu(x) e^{i\phi_\mu(x)}\right]
     + \text{rest} \,.
\end{align}
Next, we perform the eigenvalue decomposition of the transformed link, 
and rewrite the action as
\begin{align}
    S_\text{W}[U] &= 
     -\frac{\beta}{\Nc} 
     \mathbb{ReTr} \left[\Omega_\mu(x) \Lambda_\mu(x) \Omega_\mu^\dagger(x) S_\mu(x) 
      e^{i\phi_\mu(x)}\right]
      + \text{rest} \,.
     \label{eq:WilsonAction:fully-decomposed}
\end{align}
The building blocks in equation \eqref{eq:WilsonAction:fully-decomposed} are 
completely gauge invariant.%
\footnote{Note that the modal matrix $\Omega_\mu(x)$ is not unique unless one fixes 
the algorithm used for the eigenvalue decomposition.}
Thus, any transformation of these components is gauge-equivariant. 
We proceed by transforming both the \emph{spectral} and \emph{modal} matrices, 
namely $\Lambda$ and $\Omega$ to $\Lambda'$ and $\Omega'$, as discussed below. 
Afterward, we construct the updated values of $\tilde U_\mu(x)$ and $U_\mu(x)$ as
\begin{align}
   \tilde{U}'_\mu(x) &= Q'_\mu(x) \Lambda'_\mu(x) {Q'}_\mu^\dagger(x) , \\
   U'_\mu(x) &= V_\mu(x) \tilde{U}'_\mu(x) W_\mu^\dagger(x) e^{i\phi_\mu(x)}\,.
\end{align}
The advantage of this method, compared to the plaquette-based spectral flow
described above,
is that we update the link variables by incorporating information 
from all adjacent plaquettes. Moreover, we use masks that divide all plaquette Wilson loops into active and frozen ones --- there are no passive loops.

For \sutw~gauge theories, the proposed decomposition is particularly convenient 
because the phase $\phi_\mu(x)$ vanishes, the matrix of singular values is proportional 
to the identity matrix, and the spectral matrix of the transformed link
becomes a diagonal matrix consisting of a phase factor and its conjugate:
\begin{align}
 S_\mu(x) &= \sigma_\mu(x) 
 \left( \begin{array}{cc} 1 & 0 \\ 0 & 1 \end{array} \right) , \quad
 \Lambda_\mu(x) = \left(\begin{array}{cc} e^{ i\theta_\mu(x)} & 0 \\ 0 & e^{-i\theta_\mu(x)} \end{array}\right) .
\end{align}
Putting things together, the \sutw~Wilson action reads
\begin{align}
    S_\text{W}[U] \Big|_{\mathrm{SU}(2)} &= 
     -\frac{\beta}{\Nc} \sigma_\mu(x) \cos(\theta_\mu(x))
     + \text{rest} \,.
     \label{eq:WilsonAction:SU2}
\end{align}

\begin{figure}
    \begin{center}
    \includegraphics[width=1.9in]{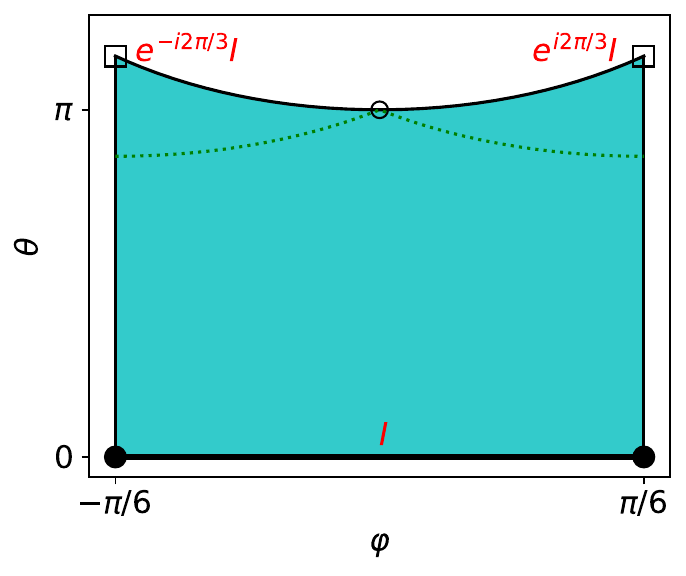}
    \includegraphics[width=3.2in]{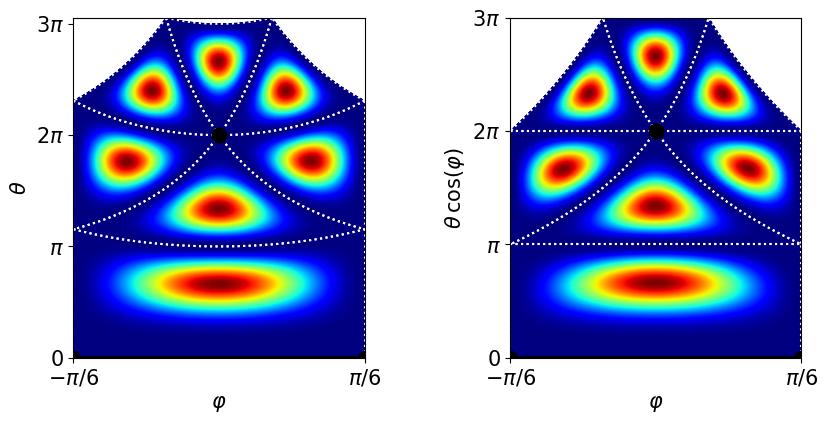}
    \end{center}
    \vspace{-0.5cm}
    \caption{Left panel shows the principal cell of the parametrization introduced in
    Eq.~\eqref{eq:SU3:eig:parametrization}. Middle and right panels show more cells and conjugacy volume as a color coding. The dotted white lines separate different cells.}
    \label{fig:SU3:cells}
\end{figure}

For \sut~gauge theories, we use the following parametrization of the spectral
matrix~\cite{Curtright:2015_SU3}:
\begin{align}
    \label{eq:SU3:eig:parametrization}
    \Lambda_\mu(x) 
    = \left(
    \begin{array}{ccc}
    e^{i \theta s_1} & 0 & 0 \\
    0 & e^{i \theta s_2} & 0 \\
    0 & 0 & e^{i \theta s_3}
    \end{array}
    \right),
    & \quad
    s_k = \frac{2}{\sqrt{3}} \sin \left(\varphi + \frac{2\pi}{3} k\right),
    \quad (s_1 + s_2 + s_3 = 0) \nonumber.
\end{align}
The parametrization of the eigenvalues in terms of $(\theta, \varphi)$ is not 
unique. The left panel of Fig.~\ref{fig:SU3:cells} shows the principal cell
of this parametrization, where the vertical and horizontal axes represent $\theta$ 
and $\varphi$, respectively. The middle panel displays several equivalent cells 
separated by dotted white lines. In the right panel, the vertical axis is replaced 
by $\theta \cos(\varphi)$, making the principal cell rectangular. The color coding 
in the middle and right panels indicates the conjugacy volume of the eigenvalues.

We parameterize the eigenvalues of \sut~matrices as $\left(e^{i\theta_1}, e^{i\theta_2}, e^{i\theta_3}\right)$,  
where $e^{i\theta_k}$, $k=1,2,3$, are the three eigenvalues, $-\pi \leq \theta_k < \pi$ and $\theta_1 \leq \theta_2 \leq \theta_3$. 
The sum of these phases can be either 0, $-2\pi$, or $2\pi$. In the second and  
third cases, we adjust the phases by adding $2\pi$ and $-2\pi$ to the first  
and last phases, respectively.
As a consequence of these adjustments, the sum of the phases is always 0.
We sort these phases and label them as $(x, y, z)$.
Next, we identify $(x, y, z)$ with $(\theta s_1, \theta s_2, \theta s_3)$ 
and map $(x, y, z)$ to $(\theta, \varphi)$ using
\begin{align}
  \theta \cos(\varphi) = z - x \,,\quad & \quad
  \tan(\varphi) = \frac{y}{z - x}\,.
\end{align}
We then employ rational quadratic splines
(RQS)~\cite{Gregory:1982rqs, Delbourgo:1983rqs, Durkan:2019rqsf}
to transform $\theta \cos(\varphi)$ and $\tan(\varphi)$. This leads to the
transformation of the spectral matrix $\Lambda_\mu(x)$ to $\Lambda'_\mu(x)$.
In our implementation, the transformations depend on the singular values too.
Furthermore, we transform the modal matrix; the details of this transformation will be discussed in the follow-ups to this work.

\section{Simulation results for \sut~gauge theory}
\label{sec:simulations}

The SVD-based transformation introduced above can be executed efficiently in
parallel by employing a mask that ensures the invertibility of the
transformation. For each direction $\mu$, we transform the gauge link variable
$U_\mu(x)$ at all sites $x$ in two sequential steps: updating the even  
sites followed by the odd sites.
Consequently, each transformation block that operates on all link variables
is composed of $2d$ sub-blocks on a $d$-dimensional lattice.
As a basic representative model, we construct one transformation block for
a four-dimensional lattice, containing 1288 parameters.

\begin{figure}
    \begin{center}
    \includegraphics[trim=0 0 0 1cm, clip, width=0.7\textwidth]{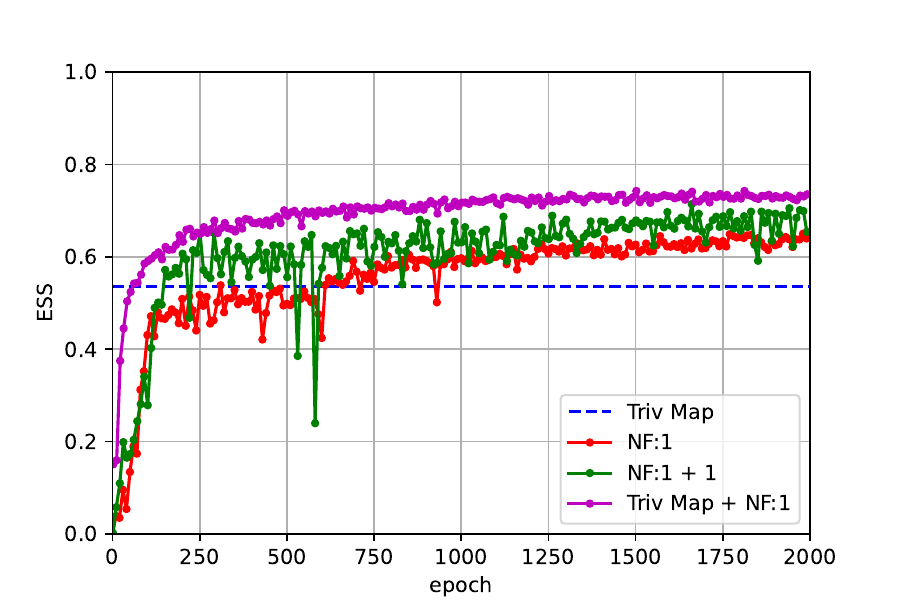}
    \end{center}
    \vspace{-0.5cm}
    \caption{The ESS, as defined in Eq.~\eqref{eq:ess:definition}, as a function of the epoch number is shown for various models applied to a lattice of size \(4^4\) using the Wilson  
    action for \sut~gauge links with \(\beta = 1\). The ESS values are measured every  
    10 epochs, with batch sizes set to 16364, 8192, 1024, and 8192 for the models  
    labeled as \text{``Triv Map''}, \text{``NF:1''}, \text{``NF:1+1''}, and  
    \text{``Triv Map + NF:1''}, respectively.
    The larger fluctuations in the ``NF:1+1'' data points (green circles) are due to the smaller
    batch size used in that case. This plot can be directly compared to Fig.~6  
    in \cite{Abbott:2023thq}.
    }
    \label{fig:training:results}
\end{figure}

We now discuss the training of our SVD-based models.
For training, we use the \emph{path-gradient estimator}~\cite{vaitl:2022pge} to  
compute the derivative of the KL divergence. Training is restricted to 2000 epochs  
with a batch size of 64.
To measure training efficiency, we use the effective sample  
size (ESS):  
\begin{equation}  
    \text{ESS} = \left.  
    {\left(\mathbb{E}_q \frac{p[U]}{q[U]} \right)^2}  
    \right /  
    {\mathbb{E}_q\left(\frac{p[U]}{q[U]}\right)^2}\, .  
    \label{eq:ess:definition}  
\end{equation}
Figure~\ref{fig:training:results} shows the evolution of the ESS for three \sut~models
on a $4^4$ lattice with $\beta = 1$.
For comparison, the leading-order trivializing map is shown with a dashed line.
This map has no trainable parameters, and its ESS is approximately $1/2$.
The training progress for our SVD-based representative model with 1288 parameters
is shown with red points.
The model's ESS matches the reference point after about $500$ epochs
and surpasses it as training progresses.
Cascading two such blocks creates a larger model,  
generally outperforming the initial model (green points). Additionally, we trained  
a model combining the leading-order trivializing map with one block layer. As  
shown by magenta points, this combination improves efficiency in terms of training
epochs.

Our investigations show that our SVD-based representative models train
significantly better in four dimensions than a plaquette-based spectral flow with
a similar number of parameters.
In fact, we could not train any of such models to yield ESS of order 0.1
or larger (with number of parameters of size of a few thousands).
In contrast, in two dimensions, the latter generally exhibits slightly better
performance. This difference can be attributed to the  
distinct dynamics of how these models update the link variables based on the
plaquette Wilson loops. Specifically, in the plaquette-based spectral flows,
\(2d - 3\) plaquettes are passively updated for each link.
While only one plaquette per link needs to be updated in two dimensions, it increases to five in four dimensions, reducing the
competitiveness of the plaquette-based spectral flows relative to the SVD-based
approach.

\section{Summary and concluding remarks}
\label{sec:conclusion}

In this work, we investigated the application of normalizing flows to generate  
lattice gauge configurations, with a particular focus on the \sut~Wilson action.
By introducing a novel SVD-based approach to construct gauge-equivariant
transformations, we addressed the challenge of preserving gauge symmetry
directly within the transformation process while incorporating information
from all plaquette Wilson loops sharing the same link variables.
Our representative model demonstrated the effectiveness of this approach
for four-dimensional lattices, as reflected in the training process.

In the representative model, we employed element-wise maps to update the
\emph{transformed} links $\tilde U_\mu(x)$.
To further enhance efficiency, future work could explore more complex transformations,
such as coupling layers~\cite{Papamakarios:2021abc, Albergo:2021vyo, DelDebbio:2021qwf, Komijani:2023fzy}, which enable richer dependencies across sites.

In conclusion, the proposed SVD-based approach provides a promising framework for  
improving lattice gauge theory simulations through neural network-based methods.  
Incorporating gauge symmetry directly into transformations not only simplifies  
training but also ensures consistent symmetry preservation from prior to target distribution,
eliminating unintended gauge fixing. Future directions  
include deploying heavier neural network architectures, as well as investigating larger volumes
and more complex systems.

\clearpage
\bibliographystyle{apsrev4-1}
\bibliography{references.bib}

\end{document}